\documentclass{llncs}

\usepackage{amsmath}
\usepackage{amssymb}
\usepackage{amsthm}
\usepackage{amsfonts}
\usepackage{bbm}
\usepackage{bm} 
\usepackage{etoolbox}
\usepackage{float}
\usepackage{graphicx}
\usepackage{hyperref}
\usepackage{indentfirst}
\usepackage{mathrsfs}
\usepackage{multirow}
\usepackage{verbatim}
\usepackage{yfonts}

\theoremstyle{definition}

\newtheorem*{notation}{Notation}

\theoremstyle{remark}

\theoremstyle{remark}

\newcommand{\rk}{\text{rk}}

\newcommand{\F}{\mathbb{F}}

\newcommand{\RQCSD}{{\sf RQCSD}}

\newcommand{\pk}{{\sf pk}}
\newcommand{\sk}{{\sf sk}}
\newcommand{\Supp}{{\sf Supp}}
\newcommand{\rot}{\text{rot}}

\begin{document}

\title{Key Recovery Attack on Rank Quasi-Cyclic Code-based Signature Scheme}

\author{Terry Shue Chien Lau \and Chik How Tan}
\institute{Temasek Laboratories, \\
National University of Singapore, \\
5A Engineering Drive 1, \#09-02, \\
Singapore 117411 \\
\email{\{tsltlsc,tsltch\}@nus.edu.sg}}

\maketitle              

\begin{abstract}
Rank Quasi-Cyclic Signature (RQCS) is a rank metric code-based signature scheme based on the Rank Quasi-Cyclic Syndrome Decoding
($\RQCSD$) problem proposed by Song et al. in \cite{SHMW19}. Their paper was accepted in the 22nd International Conference on Practice and Theory of Public Key Cryptography (PKC 2019). They have also shown that RQCS is {\sf EUF-CMA} in the random oracle model. This short paper describes how to recover the secret key in RQCS with practical simulations. Our experimental results show that we are able to recover the secret key of RQCS in less than 41 seconds for all the proposed schemes at $128$-bit, $192$-bit and $256$-bit security level.

\end{abstract}

\section{Introduction} \label{section1}

\noindent Rank Quasi-Cyclic Signature (RQCS) is a rank metric code-based signature scheme  based on the Rank Quasi-Cyclic Syndrome Decoding ($\RQCSD$) problem proposed by Song et al. in \cite{SHMW19}. Recently, their paper was accepted in the 22nd International Conference on Practice and Theory of Public Key Cryptography (PKC 2019). They have also shown that RQCS is {\sf EUF-CMA} in the random oracle model.

This short paper describes how to recover the secret key $(\bm{x},\bm{y})$ in RQCS. From a signature $(\bm{g},\bm{u})$, we first recover a support basis for $\bm{x}$ and $\bm{y}$ from $\bm{u_1}$ and $\bm{u_2}$ respectively. Then, we recover a support matrix for $\bm{x}$ and $\bm{y}$ from the public key $(\bm{h},\bm{s})$. We show the result of the simulations of our attack at the end of paper. The full version of our paper will be made available later.

\begin{notation}
  Denote the rank weight of a vector as $|| \cdot ||$ .
\end{notation}

\section{RQCS Signature Scheme} \label{background}

\noindent Let $\mathcal{H} : \mathcal{S}_{w_r} \times \{ 0,1\}^* \rightarrow \mathcal{E}_{w_g} $ be a collision-resistant hash function, where
\begin{align*}
  \mathcal{S}_{w_r} &= \{ \bm{e_1} + \bm{h} \cdot \bm{e_2} \mid \bm{e}=(\bm{e_1},\bm{e_2}) \in {\F_{q^m}^{2n}} \text{ st. } || \bm{e_1} || = || \bm{e_2} || = w_r  \}. \\
  \mathcal{E}_{w_g} &= \{ \bm{g} \in {\F_{q^m}^n} \mid || \bm{g} || = w_g \}.
\end{align*}

\noindent Let $\bm{a},\bm{b},\bm{c} \in \F_{q^m}^n$ and $(\cdot):\F_{q^m}^n \times \F_{q^m}^n \to \F_{q^m}^n$ be the rotational product between two vectors, satisfying
\[ \bm{a} \cdot \bm{b} = \bm{a} \left[ \rot(\bm{b}) \right]^T = \left[ \rot(\bm{a}) \bm{b}^T \right]^T = \bm{b} \left[ \rot(\bm{a}) \right]^T   = \bm{b} \cdot \bm{a} \]
where $\left[ \rot(\bm{b}) \right]^T$ is the circulant matrix generated by $\bm{b}$. By abuse of notation, Song et al. \cite{SHMW19} define $(\cdot): \F_{q^m}^{2n} \times \F_{q^m}^n \to \F_{q^m}^{2n}$ as $(\bm{a},\bm{b}) \cdot \bm{c} = (\bm{a}\cdot\bm{c},\bm{b} \cdot \bm{c})$. 

\medskip

\noindent Now we describe the RQCS signature scheme.

\medskip

\noindent \textbf{\sf RQCS.Setup}: Taking the security parameter $1^\lambda$ as input, it generates the public parameters ${\sf param} = (n,w,w_r,w_g)$.

\medskip

\noindent \textbf{\sf RQCS.Gen}: Taking ${\sf param}$ as input, it chooses a uniform $\bm{h} \in {\F_{q^m}^{n}}$ and $(\bm{x},\bm{y}) \in {\F_{q^m}^{2n}}$ such that $||x|| = ||y|| = w$. It computes $\bm{s} = \bm{x} + \bm{h} \cdot \bm{y}$ and outputs a pair of keys $(\pk,\sk)$. The public key $\pk$ is $(\bm{h},\bm{s})$ and the private key $\sk$ is $(\bm{x},\bm{y})$.

\medskip

\noindent \textbf{\sf RQCS.Sign}: Taking a private key $\sk = (\bm{x},\bm{y})$ and a message $\bm{m}$ as input, it chooses a uniform $\bm{r} = (\bm{r_1},\bm{r_2}) \in {\F_{q^m}^{2n}}$ such that $||\bm{r_1}|| = ||\bm{r_2}|| = w_r$. computes $I = \bm{r_1} + \bm{h} \cdot \bm{r_2}$ and $\bm{g} = \mathcal{H} (I,\bm{m})$ with $|| \bm{g} || = w_g$, followed by $\bm{u}=(\bm{u_1},\bm{u_2})=(\bm{x},\bm{y}) \cdot \bm{g} + \bm{r}=(\bm{x}\cdot \bm{g},\bm{y}\cdot \bm{g})  + (\bm{r_1},\bm{r_2})$. Then outputs the signature $(\bm{g},\bm{u})$.

\medskip

\noindent \textbf{\sf RQCS.Vrfy}: Taking a public key $\pk = (\bm{h},\bm{s})$, a message $\bm{m}$, and a signature $(\bm{g}, \bm{u})$ as input. It computes $I = \bm{u_1} + \bm{h} \cdot \bm{u_2} - \bm{s} \cdot \bm{g}$ and outputs $1$ if and only if $\mathcal{H}(I,\bm{m}) = \bm{g}$, $|| \bm{u_1} || \leq ww_g+w_r$, and $|| \bm{u_2} || \leq ww_g+w_r$.

\section{Our Key Recovery Attack on RQCS} \label{KRA}

\noindent Recall that for a vector $\bm{e} = (e_1,\ldots,e_n) \in {\F_{q^m}^n}$ with $|| \bm{e} || = w$, there exists a vector $\bm{\hat{e}} = (e_1,\ldots,e_w) \in \F_{q^m}^w$ with $|| \bm{\hat{e}}|| = w$ and a $w \times n$ matrix $E$ over $\F_q$ with $\rk (E) = w$ such that $\bm{e} = \bm{\hat{e}}E$. Note that $\bm{\hat{e}}$ and $E$ are non unique. We call $\bm{\hat{e}}$ and $E$ satisfying $\bm{e} = \bm{\hat{e}}E$ as a support basis and a support matrix for $\bm{e}$ respectively. We denote the support space of $\bm{e}$, $\Supp(\bm{e}) = \langle e_1,\ldots,e_n \rangle$ as the subspace of ${\F_{q^m}}$ generated by the coordinates of $\bm{e}$.

\medskip

We extend the idea of \cite[Algorithm 1]{GRSZ14} to solve for a support basis for $\bm{x}$. Given $\bm{a} = \bm{b}C \in \F_{q^m}^{n-k}$ with $\langle c_{ij} \rangle_{1\leq i \leq n, 1 \leq j \leq n-k}$ (of dimension $d$) known, the \cite[Algorithm 1]{GRSZ14} is to solve for $\bm{b} \in \F_{q^m}^n$ with $|| \bm{b} || \leq r$. In our case, we are given a vector $\bm{u_1} = \bm{x} \cdot \bm{g} + \bm{r_1} = \bm{x} \left[ \rot(\bm{g}) \right]^T + \bm{r_1}$ and a vector $\bm{g}$ with $||\bm{g}|| = w_g$, we are supposed to solve for $\bm{x}$.

There are two parts in our key recovery attack on RQCS. The first part is to recover a support basis for the vector $\bm{x}$ and $\bm{y}$ from $\bm{u_1}$ and $\bm{u_2}$ respectively. The second part is to recover a support matrix for $\bm{x}$ and $\bm{y}$ from $\pk = (\bm{h},\bm{s})$. Once we have recovered a support basis and a support matrix for $\bm{x}$ and $\bm{y}$, we can then recover the secret key $\sk = (\bm{x}, \bm{y})$.

\medskip

\noindent \textbf{Step 1 - Recover a Support Basis for $\bm{x}$ and $\bm{y}$.}

\medskip

\noindent Let $\mathcal{F}_x = \Supp (\bm{x})$, $\mathcal{F}_y = \Supp (\bm{y})$ and $\mathcal{G} = \Supp (\bm{g}) = \langle \gamma_1,\ldots,\gamma_{w_g} \rangle$. We first discuss the steps to recover a support basis for $\bm{x}$ from $\bm{u_1} = (u_{11},\ldots,u_{1n}) = \bm{x} \cdot \bm{g} + \bm{r_1}$:
\begin{enumerate}
  \item[(a)] For $1 \leq i \leq w_g$, compute $U_{1i} = \langle \gamma_i^{-1} u_{11}, \ldots, \gamma_i^{-1} u_{1n} \rangle$.
  \item[(b)] Compute $\mathcal{F}_x = \bigcap_{i=1}^{w_g} U_{1i} $.
  \item[(c)] Compute a basis $\{ \alpha_1,\ldots,\alpha_w\}$ for $\bm{x}$ from $\mathcal{F}_x = \langle \alpha_1,\ldots,\alpha_w \rangle$.
\end{enumerate}
\noindent Similarly, we can repeat step 1(a) to 1(c) on $\bm{u_2} = \bm{y} \cdot \bm{g} + \bm{r_2}$ to recover a basis $\{ \beta_1,\ldots,\beta_w \}$ for $\bm{y}$.

\medskip

\noindent \textbf{Step 2 - Recover Support Matrices for $\bm{x}$ and $\bm{y}$.}

\medskip

\noindent Recall that $\bm{h} \cdot \bm{y} = \bm{y} \cdot \bm{h} = \bm{y} \left[ \rot(\bm{h}) \right]^T$. Now, let $\bm{\alpha} = (\alpha_1,\ldots,\alpha_w)$ and $\bm{\beta} = ( \beta_1,\ldots,\beta_w )$ computed from step 1. Also, let $R = \left[ \rot(\bm{h}) \right]^T$.

We know that there exists $X,Y \in \F_q^{w \times n}$ with $\rk (X) = \rk(Y) = w$ such that $\bm{x} = \bm{\alpha}X$ and $\bm{y} = \bm{\beta}Y$. Substituting these two equations into $\bm{s}$:
\begin{align}
  \bm{s} &= \bm{x} + \bm{h} \cdot \bm{y} = \bm{x}  + \bm{y} R = \bm{\alpha}X + \bm{\beta}YR. \label{keyeqn}
\end{align}
The linear system (\ref{keyeqn}) consists of $n$ equations over $\F_{q^m}$ with a total of $2wn$ unknown variables (from $X$ and $Y$) to be solved. Consider (\ref{keyeqn}) under the base field $\F_q$, we have $nm$ equations over $\F_q$ with a total of $2wn$ unknown variables. Since the inequality $m>2w$ is always true in the construction of RQCS, we have $nm > 2wn$, the number of equations is more than the number of unknown variables. Then we are able to recover the matrices $X$ and $Y$ in polynomial time.

\medskip

\noindent \textbf{Simulations of Our Attack on RQCS.} We consider all the parameters of RQCS given in \cite{SHMW19} and perform simulations of our key recovery attack. The experimental results of our key recovery attack are presented in Table \ref{results}. The experiments were performed using Magma V2.20-5 running on a 3.4 GHz Intel(R) Core$^{\text{TM}}$ i7 processor with 16GB of memory.

We experimented with all the three proposed parameters: RQCS-1, RQCS-2, and RQCS-3. For each parameter, we measured the time taken (denoted as ``KRA Time'') to recover the secret key with our algorithm. The probability of failure for our algorithm is denoted as ``$p_f$''. Table \ref{results} presents the average timing of 100 experiments for each parameter.

\begin{table}[H]
  \centering
  \begin{tabular}{c|c|c|c}
  \hline
  \, Instances \, & \, $(q,m,n,w=w_r=w_g)$ \, & \, Claimed Security \, & \, KRA Time \, \\
  \hline
  RQCS-1 & $(2,89,67,5)$ & 128 & 8.17 seconds  \\
  RQCS-2 & $(2,121,97,6)$ & 192 & 32.13 seconds  \\
  RQCS-3 & $(2,139,101,6)$ & 256 & \, 40.89 seconds \,\\
  \hline
  \end{tabular}
  
  \vspace{1mm}  
  
  \caption{Simulations results of our key recovery attack against RQCS} \label{results}
\end{table}

\noindent Our key recovery attack is able to recover the secret key of all the RQCS schemes on an average time of less than $41$ seconds. 

\section{Concluding Remark}

\noindent We have proposed a key recovery attack to recover the secret key $\sk = (\bm{x},\bm{y})$ for RQCS signature scheme. In fact, our attack can be considered as a full cryptanalysis of the RQCS, in the sense that we do not attack only the parameters, but also attack the structure of the system. More specifically, we can always determine a support basis for the secret key $\bm{x}$ and $\bm{y}$, due to the properties that $|| \bm{u_1} || \leq ww_g+w_r$ and $|| \bm{u_2} || \leq ww_g+w_r$ with $\bm{g}$ publicly known. Furthermore, to verify the rank $|| \bm{u_1} || \leq ww_g+w_r$ and $|| \bm{u_2} || \leq ww_g+w_r$, it is required that $w w_g + w_r < \min \{ m,n\}$, which ensures the conditions of $m > 2w$ for us to solve for support matrices of $\bm{x}$ and $\bm{y}$. In conclusion, RQCS is completely cryptanalyzed by our key recovery attack. Our key recovery attack is able to recover the secret key of all the RQCS schemes on an average time of less than $41$ seconds.

\end{document}